\documentclass{iopart}
\usepackage{graphics,graphicx}
\usepackage{subfigure}

\usepackage{iopams}

\def \bsigma {\mbox {\boldmath $\sigma$}}
\def \bxi {\mbox {\boldmath $\xi$}}
\def \bS {\mbox {\boldmath $S$}}
\def \bpsi {\mbox {\boldmath $\psi$}}
\def \bphi {\mbox {\boldmath $\phi$}}
\def \bC {\mbox {\boldmath $C$}}
\def \bG {\mbox {\boldmath $G$}}

\def \bI {\mbox {\boldmath $I$}}
\def \bA {\mbox {\boldmath $A$}}
\def \bB {\mbox {\boldmath $B$}}
\def \bV {\mbox {\boldmath $V$}}
\def \bU {\mbox {\boldmath $U$}}
\def\ket#1{\mathinner{|{#1}\rangle}}

\begin{document}

\title{Instability of frozen-in states in synchronous Hebbian neural
  networks}
\author{F L Metz and W K Theumann}
\address{Instituto de F\'\i sica, Universidade Federal do Rio
Grande do Sul, Caixa Postal 15051, 91501-970 Porto Alegre, Brazil}
\eads{\mailto{theumann@if.ufrgs.br}, \mailto{fmetz@if.ufrgs.br}}

\date{\today}
\thispagestyle{empty}

\begin{abstract}

The full dynamics of a synchronous recurrent neural network model
with Ising binary units and a Hebbian learning rule with a finite
self-interaction is studied in order to determine the stability to
synaptic and stochastic noise of frozen-in states that appear in
the absence of both kinds of noise. Both, the numerical simulation
procedure of Eissfeller and Opper and a new alternative procedure
that allows to follow the dynamics over larger time scales have
been used in this work. It is shown that synaptic noise
destabilizes the frozen-in states and yields either retrieval or
paramagnetic states for not too large stochastic noise. The
indications are that the same results may follow in the absence of
synaptic noise, for low stochastic noise.

\end{abstract}

\submitto{\JPA}


\section{Introduction}

The dynamics of recurrent neural network models with synchronous
updating has been of interest ever since the work of Little
\cite{Li74,FK87,AC01}. It is expected that Little's model with
binary units and symmetric Hebbian couplings should exhibit either
cycles of period two or fixed-point attractors and it has been
suggested that the cycles could arise in statistical mechanics from
a duplication of phase space by means of two state variables for
every unit. An equilibrium analysis, based on replica symmetry,
predicted that period-two cycles associated with a full spin flip
(all units changing sign at each time step) should occur in a zero
temperature paramagnetic phase. This was recognized as an unphysical
phase apparently disconnected from the ordinary high temperature
paramagnetic phase. A somewhat large negative self-interaction $J_0$
(the diagonal elements of the synaptic matrix) between units turned
out to be necessary for this cyclic solution with full spin flip to
appear \cite{FK87}.

Early numerical simulations failed to show the presence of those
cycles and lead to the conjecture that the problem with the
statistical mechanics approach could be due to the assumption of
replica symmetry. Indeed, the entropy of this phase was found to
be negative, going to $-\infty$ as $T \rightarrow 0$ \cite{FK87},
a feature that usually goes together with replica symmetry
\cite{PMV87}. In a further work, a zero-temperature calculation of
the average number of cycles as a function of the fraction of
flipping spins demonstrated that period-two cycles with full spin
flip are by far the most common type of cycles below a storage
ratio of patterns $\alpha \approx 0.7$ \cite{Fo97}. However, the
usual calculation of the average number of metastable states does
not say anything about the macroscopic properties, in particular
about the overlap with a chosen pattern.

Fixed-point solutions for the macroscopic parameters that emerge
from a dynamics with a finite fraction of spins changing sign at
each time step are common, but not exclusive, to the phase diagram
of Little's model. Indeed, they also appear in recent studies of
three-state Ising and Blume-Emery-Griffiths neural network models
with synchronous updating. These models can be thought as
extensions of Little's model with generalized synaptic
interactions and multi-state units \cite{Bo04}. Only a small
fraction of neurons that change sign appear to be involved in the
stationary states \cite{BB05,BEV06}, and the work in those studies
was mainly devoted to fixed-point solutions for the macroscopic
parameters. On the other hand, stationary period-two solutions for
the macroscopic parameters, that arise from a fraction of units
changing sign at each time step, have been found in recent work on
the synchronous dynamics of symmetric sequence processing without
a self-interaction \cite{MT07}. The relation between the dynamics
of the macroscopic parameters and the fraction of units that
change sign at each time step is crucial to understand the
behavior of synchronous networks.

A dynamical study may be helpful to get further insight into the
asymptotic behavior of Little's model. So far, only the results of
an approximate synchronous dynamics are available, which become
exact for an asymmetrically random diluted network in the limit of
extreme dilution \cite{Fo88}. This dynamics predicts that, at
$T=0$ and $\alpha=0$, either frozen-in fixed points or frozen-in
cycles of period two may appear when $|J_0|>|m_0|$, for a positive
or a negative self-interaction, respectively, where $m_0$ is the
initial overlap with a chosen pattern. On the other hand, there is
a flow to a fixed point $m_t=1$ reached in time $t$, when
$|J_0|<|m_0|$. Thus, both the size and the sign of the
self-interaction play a crucial role and the same dynamics, for
small non-zero $\alpha$, predicts a flow either to a paramagnetic
or to a retrieval state.

Frozen-in states do not evolve in time and they are an undesirable
feature for associative memory. In the present context, they are
states in which the overlap either remains fixed at $m_0$
(frozen-in fixed-point) or oscillates between $m_0$ and $-m_0$
(frozen-in cycle of period two). The correlation function between
two consecutive time steps, related to the fraction of flipping
spins, is $C_{t,t-1}=1$ for a frozen-in fixed-point and
$C_{t,t-1}=-1$ for a frozen-in cycle of period two. The question
that arises is if the frozen-in states actually are stable in
Little's fully connected model subject to synaptic and/or
stochastic noise, and the main purpose of this paper is to deal
with that issue following the transient behavior of the dynamics
that yields the stationary states of the network. The full
dynamics of the Hebbian synchronous recurrent network of binary
units, either at zero or at finite temperature with stochastic
noise, including eventually the region where period-two cycles
dominate has, apparently, not been carried out before. We make use
of a generating functional approach (GFA) \cite{DD78,AC01II},
exact in the mean-field limit of an infinitely large system,
combined with the numerical simulation procedure of Eissfeller and
Opper (EO), based on the GFA for the dynamics of disordered spin
systems. This is a procedure free of finite size effects
\cite{EO92,Ve03}.

The problem with spin-glass-like models (neural networks among them)
is that the dynamics may be very slow. Due to that, in particular
for not too small $|J_0|$ in both cyclic and in retrieval states,
the numerical simulation may require a large number of time steps in
order to reach a stationary behavior, making the computation
prohibitive with the EO procedure. A further purpose of the paper is
to overcome this problem by means of an alternative procedure that
is introduced in this work and which consists mainly of two
features. One is the neglect of memory effects that involve the
response function in the self-energy term of the single-site
effective field, which amounts to a signal-to-noise approximation
(SNA) \cite{BBV04}. The second feature consists in doing
analytically the summation over states in the GFA. Thereby one
accounts for the microscopic variables in the calculation of the
dynamics of the macroscopic properties in a way that involves a much
smaller number of variables. The procedure is exact for $\alpha=0$,
and yields an approximation for small $\alpha$, that can be done in
a short computing time up to a large number of time steps producing
results in excellent agreement with the procedure of EO for finite
$T$ and sufficiently large values of $|J_{0}|$. We make use of both
our and the EO procedure in what follows.

The outline of the paper is the following. In Section 2 we
introduce the model and present a brief summary of the now well
known GFA supplemented with an adaptation of the EO procedure to
our system. In Section 3 we discuss our alternative procedure and
in Section 4 we show the results of the transient dynamics up to
the asymptotic states for the overlap and the correlation function
between two consecutive times. We conclude with a summary and a
further discussion in Section 5.

\section{The model and the generating functional approach}

We consider a network of $N$ Ising neurons in a microscopic state
$\bsigma(t)=\{\sigma_1(t),\dots,\sigma_N(t)\}$, at the time step $t$
in which each $\sigma_i(t)=\pm 1$ represents an active or inactive
neuron, respectively. The states of all neurons are updated
simultaneously at each discrete time step according to the alignment
of each spin with its local field
\begin{equation}
h_i(t)=\sum_{j}J_{ij}\sigma_j(t) + \theta_i(t)\,\,\,, \label{1}
\end{equation}
following a microscopic stochastic single spin-flip dynamics with
transition probability
\begin{equation}
w[\sigma_i(t+1)|h_i(t)] =\frac{1}{2} \big{\{} 1+\sigma_i(t+1)\tanh[\beta
h_i(t)] \big{\}} \,\,\,, \label{2}
\end{equation}
ruled by the noise-control parameter $\beta=T^{-1}$, where $T$ is
the synaptic noise. The dynamics is a deterministic one when $T=0$
and fully random when $T =\infty$. In the former case,
$\sigma_i(t+1)={\rm sgn}[h_i(t)]$. Here, $\theta_i(t)$ is an
external stimulus and $J_{ij}$ is the synaptic coupling
\begin{eqnarray}
J_{ij}&=&\frac{1}{N}\sum_{\mu=1}^p\xi_i^{\mu}
\xi_j^{\mu}\,\,\,,\,\,\, i\neq j \,\,\,,\nonumber\\
&=&J_0 \,\,\,,\,\,\, i=j \,\,\,,\label{3}
\end{eqnarray}
between units $i$ and $j$, which is assumed to have a Hebbian form
when $i\neq j$, with a macroscopic set $\bxi^{\mu}
=(\xi_1^{\mu},\dots,\xi_N^{\mu})$, $\mu=1,\dots,p$ of $p=\alpha N$
independent and identically distributed quenched random patterns.
Each $\xi_i^{\mu}=\pm 1$ with probability $\frac{1}{2}$ and $J_0$ is
a variable self-interaction. The latter plays a crucial role leading
either to fixed-point or cyclic behavior.

The dynamical evolution of the system is described by the moment
generating functional \cite{AC01II}
\begin{eqnarray}
\fl Z(\bpsi) &=& \Big{\langle}  \exp\Big[-\rmi \sum_{i}
\sum_{s=0}^{t}
\psi_{i}(s) \, \sigma_{i}(s)\Big]  \Big{\rangle} \nonumber \\
\fl &=& \sum_{\bsigma(0),\dots,\bsigma(t)}
\mathrm{Prob}[\bsigma(0),\dots,\bsigma(t)] \exp\Big[-\rmi \sum_{i}
  \sum_{s=0}^{t} \psi_{i}(s) \, \sigma_{i}(s)\Big]\,\,, \label{4}
\end{eqnarray}
for a finite number of time steps $t$, where
$\bpsi(t)=(\psi_1(t),\dots,\psi_N(t))$ is a set of auxiliary variables
that generates averages of moments of the states and the brackets
denote an average over all possible paths of states with probability
\begin{equation}
\mathrm{Prob}[\bsigma(0),\dots,\bsigma(t)]
 =p[\bsigma(0)] \prod_{s=0}^{t-1} \prod_{i}
 \frac{\exp[\,\beta\,\sigma_{i}(s+1)\,
h_{i}(s)]}{2\,\cosh{[\beta\,h_{i}(s)]}} \label{5}
\end{equation}
that follows from (2). Assuming that for $N\rightarrow \infty$
only the statistical properties of the stored patterns will
influence the macroscopic behavior, one obtains the relevant
quantities which are the overlap $m_{1}(t)$ of $\Or (1)$ with the
condensed pattern $\xi^{1}$, say, the two-time correlation function
$C(t,t^{\prime})$ and the response function $G(t,t^{\prime})$, given
by
\begin{equation}
m_{1}(t) = \frac{1}{N} \sum_{i} \overline{\xi_{i}^{1} \langle
  \sigma_{i}(t) \rangle}
  = \lim_{\bpsi \rightarrow 0} \frac{\rmi}{N} \sum_{i}
  \xi_{i}^{1}  \frac{\partial \overline{Z(\bpsi)}}{\partial
  \psi_{i}(t)}\,\,,   \label{6}
\end{equation}
\begin{equation}
C(t,t^{\prime}) = \frac{1}{N} \sum_{i} \overline{\langle
  \sigma_{i}(t) \sigma_{i}(t^{\prime}) \rangle}
  =  -\lim_{\bpsi \rightarrow 0}
  \frac{1}{N} \sum_{i}  \frac{\partial^{2} \overline{Z(\bpsi)}}{\partial
  \psi_{i}(t^{\prime}) \psi_{i}(t)}\,\,, \label{7}
\end{equation}
and
\begin{equation}
G(t,t^{\prime}) = \frac{1}{N} \sum_{i} \frac{\partial
\overline{\langle \sigma_{i}(t) \rangle}}{\partial
  \theta_{i}(t^{\prime})}
  = \rmi \, \lim_{\bpsi \rightarrow 0}
  \frac{1}{N} \sum_{i}  \frac{\partial^{2} \overline{Z(\bpsi)}}{\partial
  \theta_{i}(t^{\prime}) \psi_{i}(t)}\,\,\,\, (t^{\prime} < t)\,\,,  \label{8}
\end{equation}
where the bar denotes the configurational average with
the non-condensed patters $\bxi^{\mu}$, $\mu>1$, and the restriction
$t^{\prime} < t$ is due to causality. Calling $q_0(t)=C(t,t-1)$, the
fraction of flipping spins between two consecutive times becomes
$[1-q_0(t)]/2$.

Following the now standard procedure in which the disorder average
is done before the sum over the neuron states one obtains exactly,
in the large $N$ limit, the generating functional \cite{AC01II,EO92}
\begin{eqnarray}
\fl Z(\bpsi)= \Big{\langle} \sum_{\bsigma(0),\dots,\bsigma(t)}
p[\bsigma(0)] \exp\Big{[}-\rmi\sum_i
\sum_{s=0}^{t} \psi_i(s) \, \sigma_i(s) \Big{]} \nonumber \\
 \prod_{i} \prod_{s < t}  \Big{\{} \int \rmd h_i(s)
\delta[h_i(s)-h_{i}^{\mathrm{eff}}(s)]\,
\mathrm{w}[\sigma_i(s+1)|h_i(s)] \Big{\}}
\Big{\rangle}_{\{\phi_i(s)\}} \label{9}
\end{eqnarray}
in which $p[\bsigma(0)]= \prod_{i}p[\sigma_{i}(0)]$ is the
probability of the initial microscopic configuration while $\langle
\dots \rangle_{\{\phi_i(t)\}}$ denotes an average over the
correlated Gaussian random variables $\{\phi_i(t)\}$, with
zero-average and a correlation matrix given below. These random
variables turn out to be uncorrelated on different sites and one is
left with a single-site effective theory in which a neuron evolves
in time according to the probability
\begin{equation}
 \mathrm{w}[\sigma(t+1)| h^{\mathrm{eff}}(t)]= \frac{1}{2} \Big\{  1
+ \sigma(t+1) \tanh{\big[ \beta h^{\mathrm{eff}}(t) \big] }
\Big\}\,\,, \label{10}
\end{equation}
with an effective local field given by
\begin{equation}
  h^{\mathrm{eff}}(t) = \xi m(t) + J_{0} \sigma(t)
  + \alpha \sum_{t^{\prime}<t}   R(t,t^{\prime})   \sigma(t^{\prime}) +
  \sqrt{\alpha} \phi(t)\,\,, \label{11}
\end{equation}
where we dropped the label of the condensed pattern and assumed
that $\theta(t)=0$. The two non-trivial contributions to the
effective local field for $\alpha\neq 0$ come from a retarded
self-interaction involving the matrix elements
\begin{equation}
R(t,t^{\prime}) = [\bG(\bI-\bG)^{-1}]_{t,t^{\prime}} \label{12}
\end{equation}
and the zero-average temporarily correlated Gaussian noise $\phi(t)$
with variance
\begin{equation}
S(t,t^{\prime}) = \langle \phi(t) \phi(t^{\prime})
\rangle_{\{\phi(t)\}} =
[(\bI-\bG)^{-1}\bC(\bI-\bG^{\dagger})^{-1}]_{t,t^{\prime}}\,\,.\\
\label{13}
\end{equation}
Both these terms account for memory effects in the network. The
generating functional (\ref{9}) is the functional of
Eissfeller and Opper that applies to Little's model and the
specific algorithm that implements the numerical simulations is
described in the literature \cite{EO92,Ve03}.

The dynamics of each of the macroscopic quantities, given by
(\ref{6})-(\ref{8}), is obtained from the statistics of the effective
single neuron process through the average
\begin{equation}
\langle f(\bsigma) \rangle = \int \rmd \bphi \mathrm{P}(\bphi)
  \sum_{\bsigma}  \mathrm{P}(\bsigma|\bphi) f(\bsigma) \,\,, \label{14}
\end{equation}
where $\bsigma=\{\sigma (t)\}$ and $\bphi=\{\phi (t)\}$ are now
single-site vectors that follow a path in discrete times, and
\begin{equation}
 \mathrm{P}(\bsigma|\bphi)=p[\sigma(0)] \prod_{s<t}
 \mathrm{w}[\sigma(s+1)|h^{\mathrm{eff}}(s)] \,\, \label{15} \\
\end{equation}
is the single-spin path probability given the Gaussian noise
$\bphi$ in the effective field, with a distribution
\begin{equation}
\mathrm{P}(\bphi)=\frac{1}{\sqrt{(2\pi)^{t}\det\bS}}
\exp({-\frac{1}{2} \bphi. \bS^{-1}\bphi})\,\,. \label{16}
\end{equation}

The macroscopic parameters (overlap, correlation and response
function) can, in principle, be calculated in closed form for any
time step $t$ but that becomes non-practical since it requires an
increasingly large number of macroscopic quantities. Thus, analytic
calculations are only feasible for the first few time steps and, as
it turns out, also for the asymptotic stationary state, albeit under
some conditions which are not always fulfilled.

In order to obtain the full dynamic description of the transients,
we make use of the procedure of Eissfeller and Opper in which the
effective single-site dynamics given by (\ref{10})-(\ref{13}) is simulated
by a Monte-Carlo method. There are no finite-size effects, but a
large number $N_{T}$ of stochastic trajectories has to be generated
for the single-site process and the macroscopic parameters can be
obtained from the average
\begin{equation}
\langle f(\bsigma) \rangle = \frac{1}{N_{T}} \sum_{a=1}^{N_{T}}
f(\bsigma_{a})  \,\,, \label{17}
\end{equation}
where $\bsigma_{a}$ denotes the set of spins along the path $a$. The
number of stochastic trajectories $N_{T}$ should not be confused
with the number of neurons $N$, which goes to infinity. To keep the
numerical errors small, a sufficiently large $N_{T}$ must be used.

\section{Alternative procedure}

The alternative procedure consists in doing analytically the sum
over the microscopic paths (states in discrete times) $\bsigma$,
to start with, in order to have a much smaller set of parameters
to iterate in the numerical calculations. Due to the presence of
memory effects in the terms involving the retarded
self-interaction $R(t,t^{\prime})$, given by (\ref{12}), which relate the
state of the system at time $t$ to that at all previous times, the
full elimination of the microscopic parameters becomes a
formidable task. In the following we consider the simplest
approximation that consists in assuming that $\bG=0$, which is the
so-called signal-to-noise approximation \cite{BBV04}. This
implies, in turn, that the variance of the Gaussian correlated
noise becomes $\bS=\bC$ according to (\ref{13}).

With that approximation, the effective local field with the time
step denoted as a subindex,
\begin{equation}
h_{t} = \xi m_{t} + J_{0} \sigma_{t} + \sqrt{\alpha} \phi_{t}\,\,
\label{campoefetivo1}
\end{equation}
has still a stochastic Gaussian noise for non-zero $\alpha$ and a
dependence on the spin variable $\sigma_{t}$ at the same time. We
assume an initial distribution $p(\sigma_{0}) =
\frac{1}{2}(1+\sigma_{0} \xi m_{0})$ and proceed with the summation
over states in (\ref{14}), which we denote as
\begin{equation}
g(\bphi)=\sum_{\bsigma}  \mathrm{P}(\bsigma|\bphi)
f(\bsigma)\,\,\,. \label{defg}
\end{equation}
The summation can be done by inserting the following integral
representation
\begin{equation}
  \exp({\beta \sigma_{s+1} h_{s}}) =  \int_{0}^{2 \pi} \frac{\rmd x_{s}}{\pi} \exp({\rmi
  x_{s} \sigma_{s+1}}) \cosh{(\beta h_{s} - \rmi x_{s})}  \,\,\, \nonumber
\end{equation}
into (\ref{15}), with the effective local field given by
(\ref{campoefetivo1}), in order to separate the dependence on
the states at two consecutive times. The result, substituted in
(\ref{defg}), yields
\begin{eqnarray}
\fl g(\bphi)= \int_{0}^{2 \pi} \Big{[} \prod_{s=0}^{t-1} \frac{\rmd x_{s}}{2
\pi} \Big{]} \sum_{\sigma_0 \dots \sigma_t} p(\sigma_{0}) f(\bsigma)
\exp\Big{(}{\rmi \sum_{s=0}^{t-1}
  x_s \sigma_{s+1}} \Big{)} \nonumber \\
 \prod_{s=0}^{t-1} \Big[ \cos{x_s} - \rmi \tanh{(\beta h_s)}
\sin{x_s} \Big] \,\,\,. \label{eqaux}
\end{eqnarray}
Since the argument of the exponential is linear with respect to the
set $\{ \sigma_{s} \}$, we can now do the sum over the states
separately for each time in order to obtain the form of the function
$g(\bphi)$ in the case of the overlap and the two-time correlation
function, with $f(\bsigma)$ given by $f(\bsigma) =\xi \sigma_{t}$
and $f(\bsigma) = \sigma_{t} \sigma_{s}$ ($s=0,\dots,t-1$),
respectively. The drawback is the appearance of $t$ multiple
integrals over the full set $\{ x_{s} \}$, which can ultimately be
reduced to a pair of integrals over $x_0$ and $x_{t-1}$ involving
elements of either of the matrices
\begin{equation}
  \bU = \prod_{s=1}^{t-1} \bA_{s}\,\,\,\,\,,\,\,\,\,\,\,\,
\bV = \Big( \prod_{s=1}^{t^{\prime}-1} \bA_{s}  \Big)
 \bB_{t^{\prime}} \Big( \prod_{s=t^{\prime}+1}^{t-1} \bA_{s}
\Big)\,\,,
\end{equation}
in the representation
\begin{equation}
\ket{x_{s}} = \left( \begin{array}{c}
  \cos{x_s} \\
  \sin{x_s} \end{array} \right)
\end{equation}
where
\begin{equation}
\fl \bA_{s} = \left(\begin{array}{cc} 1 & -\frac{i}{2} [
 \tanh{\beta(\xi m_{s} + J_{0} + \sqrt{\alpha} \phi_{s})} +
 \tanh{\beta(\xi m_{s} - J_{0} + \sqrt{\alpha} \phi_{s})} ] \\
0 & \frac{1}{2} [ \tanh{\beta(\xi m_{s} + J_{0} + \sqrt{\alpha}
 \phi_{s})} -  \tanh{\beta(\xi m_{s} - J_{0} + \sqrt{\alpha} \phi_{s})} ]
  \end{array} \right ) \,\,,
\end{equation}
\begin{equation}
\fl \bB_{s} = \left(\begin{array}{cc} 0 & -\frac{i}{2} [ \tanh{\beta(\xi
m_{s} + J_{0} + \sqrt{\alpha} \phi_{s})} -
  \tanh{\beta(\xi m_{s} - J_{0} + \sqrt{\alpha} \phi_{s})} ]
\\ i & \frac{1}{2} [ \tanh{\beta(\xi m_{s} + J_{0} + \sqrt{\alpha}
\phi_{s})} +  \tanh{\beta(\xi m_{s} - J_{0} + \sqrt{\alpha}
\phi_{s})} ]
  \end{array} \right ) \,\,.
\end{equation}

Performing the discrete average over the pattern $\xi$, we obtain
the explicit expressions for the macroscopic parameters, that is,
for the overlap
\begin{eqnarray}
\fl m_{t} = \Big{<} \Big[ \frac{1}{2}(1+m_{0})
 \tanh{\beta(m_{0} + J_{0} + \sqrt{\alpha} \phi_{0})} \nonumber \\
+ \frac{1}{2}(1-m_{0})  \tanh{\beta(m_{0} - J_{0} + \sqrt{\alpha}
\phi_{0})} \Big] U_{2\,2} + \rmi \, U_{1\,2} \Big{>}_{\bphi}\,\,,
\label{overlapeq}
\end{eqnarray}
for the correlation with the initial state
\begin{eqnarray}
\fl C_{t\,0} = \Big{<} \Big[ \frac{1}{2}(1+m_{0})
 \tanh{\beta(m_{0} + J_{0} + \sqrt{\alpha} \phi_{0})} \nonumber \\
- \frac{1}{2}(1-m_{0}) \tanh{\beta(m_{0} - J_{0} + \sqrt{\alpha}
\phi_{0})} \Big] U_{2\,2} + \rmi \, \,m_{0}\,U_{1\,2}
\Big{>}_{\bphi}\,\,, \label{correq}
\end{eqnarray}
and for the correlation with the states at other times $0<t^{\prime}< t$,
\begin{eqnarray}
\fl C_{t\,t^{\prime}} = \Big{<} \Big[
\frac{1}{2}(1+m_{0})
 \tanh{\beta(m_{0} + J_{0} + \sqrt{\alpha} \phi_{0})} \nonumber \\
+ \frac{1}{2}(1-m_{0}) \tanh{\beta(m_{0} - J_{0} + \sqrt{\alpha}
\phi_{0})} \Big] V_{2\,2} + \rmi \,
 V_{1\,2} \Big{>}_{\bphi}\,\,, \label{correqA}
\end{eqnarray}
where $U_{i,j}$ and $V_{i,j}$ are the elements of the matrices $\bU$
and $\bV$. Finally, for the binary spins in this work, the diagonal
elements are $C_{t,t}=1$. Here, $m_0$ and $\phi_0$ are the initial
overlap and Gaussian noise, respectively, and the full correlation
function is needed to calculate the average
\begin{equation}
  \big{<} g(\bphi) \big{>}_{\bphi} = \int \rmd \bphi \frac{\exp{\Big( -\frac{1}{2}
\bphi. \bC^{-1}\bphi} \Big) } {(2
 \pi)^{\frac{t}{2}} \sqrt{\det{\bC}}} g(\bphi)\,\,. \label{mediaphi}
\end{equation}
Note that (\ref{overlapeq})-(\ref{correqA}) are a set of coupled
equations when $\alpha \neq 0$ involving the matrix $\bC$ and the
overlap at previous time steps through the matrices $\bU$ and $\bV$.
The remaining Gaussian averages in the above equations have to be
performed numerically replacing the integral over $\bphi$ by a
discrete average over a large number $M$ of randomly generated
functions of the set of correlated Gaussian variables
$\{\phi_t^{\lambda}\}$, $\lambda=1,\dots,\, M$, at each time step
$t$, with mean zero and variance given by the correlation matrix
$\bC$.

Before presenting our numerical results we discuss briefly the case
where $\alpha=0$. Thus, in the absence of stochastic noise, the
alternative procedure becomes exact and the average over $\bphi$
drops out in (\ref{overlapeq})-(\ref{correqA}) leading to the
following two recursion relations,
\begin{equation}
\fl m_{t+1} = \frac{1}{2}(1+m_{t}) \tanh{\beta(m_{t}+J_{0})} +
\frac{1}{2}(1-m_{t}) \tanh{\beta(m_{t}-J_{0})} \,\,
\label{overlaprecurr}
\end{equation}
and
\begin{equation}
\fl C_{t+1,s} = \frac{1}{2}(C_{t,s}+m_{s})
\tanh{\beta(m_{t}+J_{0})} - \frac{1}{2}(C_{t,s}-m_{s})
\tanh{\beta(m_{t}-J_{0})}\,\, , \label{Corrrecurr}
\end{equation}
for $s<t+1$. Setting $s=t$ and noting that $C_{t,t}=1$, we get the
two-consecutive time correlation function that yields the
spin-flip order parameter $q_{0}(t)=C_{t+1,t}$ satisfying the
equation
\begin{equation}
\fl q_{0}(t)= \frac{1}{2}(1+m_{t}) \tanh{\beta(m_{t}+J_{0})} -
\frac{1}{2}(1-m_{t}) \tanh{\beta(m_{t}-J_{0})} \,\,.
\label{q_0recurr}
\end{equation}
Thus, (\ref{overlaprecurr}) and (\ref{q_0recurr}) are extensions
for all $T$ of the equations obtained before for $T=0$
\cite{Fo88}. The asymptotic stationary order parameters are given
by the fixed-point solutions of these equations, which become
\cite{FK87}
\begin{equation}
m=\frac{\sinh{(2\beta m)}}{\cosh{(2\beta m)} + \exp{(-2 \beta
J_{0})} }\,\, \label{pontofixom}
\end{equation}
and
\begin{equation}
q_0=\frac{\cosh{(2\beta m)} - \exp{(-2 \beta
J_{0})}}{\cosh{(2\beta m)} + \exp{(-2 \beta J_{0})}  }\,\,.
\label{pontofixoq_0}
\end{equation}
In the $T=0$ limit,
\begin{eqnarray}
|J_{0}| > |m_{0}| &\rightarrow& m_{t} = m_{0} \,\,\,,\,\,q_0=1\,\,\,
 \rm{if} \,\,\,\, J_{0} > 0 \nonumber \\
&\rightarrow& m_{t} = (-1)^{t} m_{0} \,\,\,,\,\,q_0=-1\,\,\,
 \rm{if} \,\,\,\, J_{0} < 0 \label{33} \\
\vspace{3cm} |J_{0}| < |m_{0}| &\rightarrow& m_{t} = \rm{sgn}(m_0)
\,\,\,,\,\,\,q_0=1\,\,\,. \nonumber
\end{eqnarray}
Thus, at $T=0$ and $\alpha=0$, the system appears in a frozen-in
state either in the initial overlap when $J_0>0$ or in a cycle
when $J_0<0$, in the first case, or else it flows to the retrieval
state in the second case. It is in the first case that we are
interested in what follows. Clearly, in that case the system is no
longer useful as an associative memory but it could again become
useful in the presence of synaptic and/or stochastic noise and it
would be interesting to see how this occurs and we discuss that
next. When the dynamics is not too slow one may employ the
procedure of EO, but otherwise we have to resort to our
alternative approximate dynamics.

\section{Results}

\begin{figure}[ht]
\center
\includegraphics[scale=0.55]{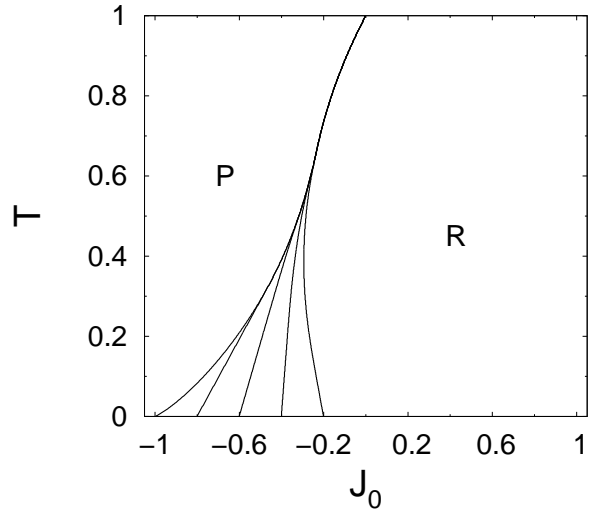}
\caption{Phase diagram for $\alpha=0$ with initial overlaps $m_{0} =
1, 0.8, 0.6, 0.4$ and $0.2$, from left to right.} \label{diagram}
\end{figure}

We start with the results for $\alpha=0$ and any $T$, for which
the alternative procedure is exact. In figure \ref{diagram} we show
the $(J_0,T)$ phase diagram of stationary states for various
initial overlaps, as indicated, obtained by means of the iteration
of (\ref{overlaprecurr}). We also analyzed the stationary
states of the correlation function between two consecutive times,
by means of (\ref{q_0recurr}). There is a paramagnetic phase
(P) to the left of the curves with $m=0$ and $q_0> -1$ for all
$T$, except at $T=0$, where there is a state of frozen-in cycles
with overlap $m_{t}=(-1)^{t}m_0$ and $q_0=-1$. The network evolves
to a retrieval fixed-point within the phase R for all $T>0$, with
an asymptotic overlap $m \simeq 1$ for low $T$ and $q_0< 1$. For
$T=0$, there is either a state of frozen-in fixed-points in phase
R with $m=m_0$ and $q_0 = 1$, if $J_0 > m_0$, or a retrieval state
with $m=1$ and $q_0 = 1$, if $J_0 < m_0$, in accordance with
(\ref{33}). The phase boundary obtained in the case of the initial
overlap $m_0=0.5$ is precisely the same as that obtained from the
equality of the free energies in the equilibrium analysis, where
the appearance of a tricritical point and other features have been
discussed \cite{FK87}.

\begin{figure}[ht]
\center
\includegraphics[scale=0.62]{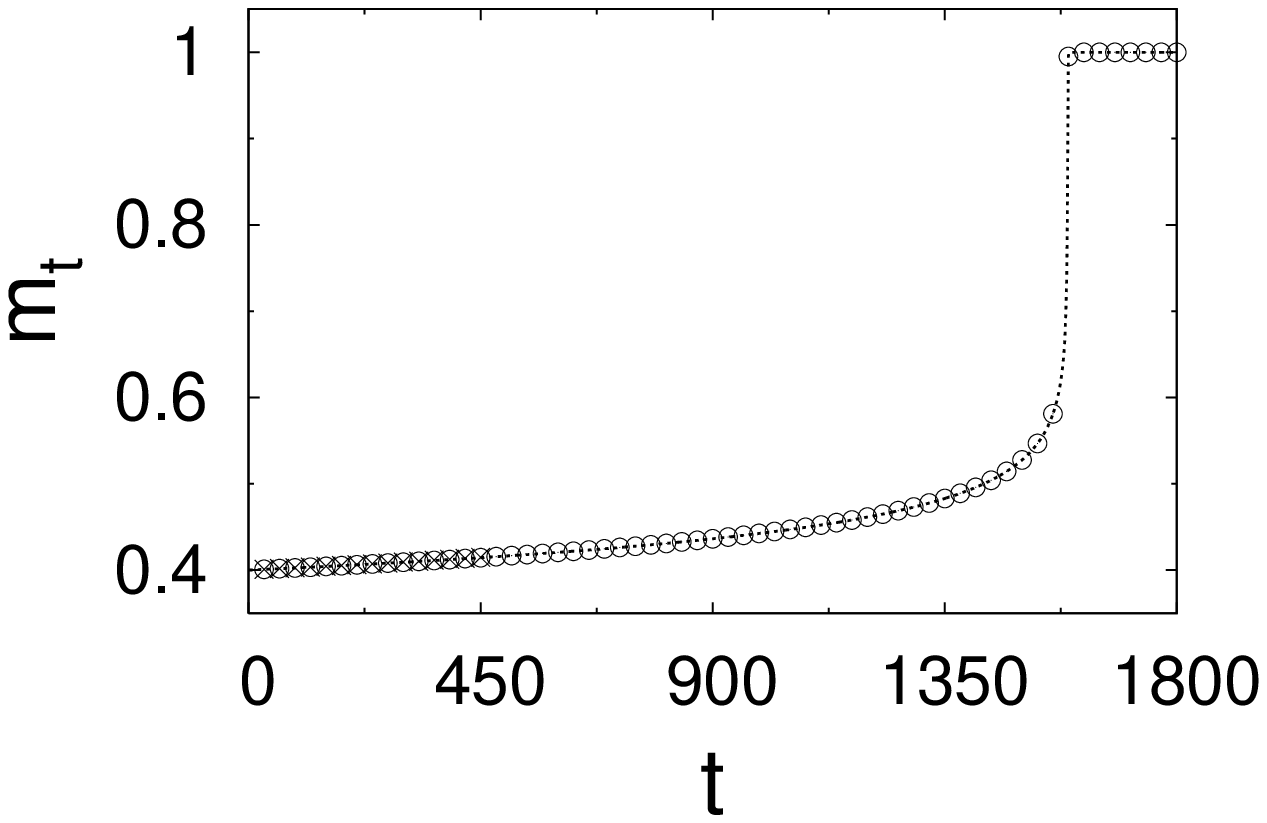}
\includegraphics[scale=0.62]{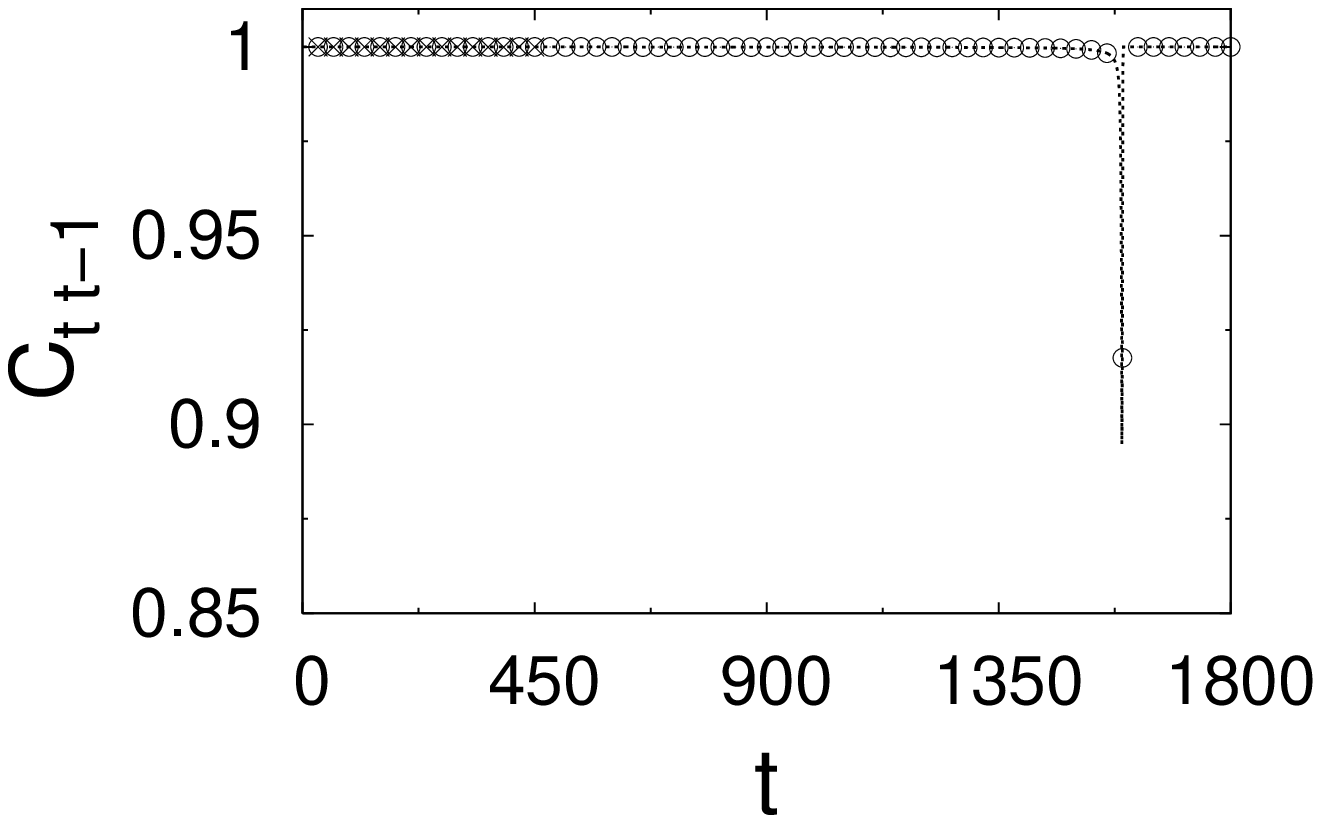}
\caption{Overlap and two-time correlation function for $T=0.08$,
$J_{0}=0.8$, $m_{0}=0.4$ and $\alpha=0$. Superimposed are the
results with the EO procedure (dark circles).}
\label{dinamalpha0retr}
\end{figure}

In order to illustrate the instability to synaptic noise of the
frozen-in fixed point in phase R, when $\alpha=0$, and the
interesting transient crossover to retrieval behavior, we show in
figure \ref{dinamalpha0retr} the dynamical evolution of the overlap
and of the two-time correlation function for $J_0=0.8$, $T=0.08$
and initial overlap $m_0=0.4$. The results were obtained by means
of the alternative procedure and, for comparison, we also show the
results with the EO procedure. For these values of the parameters,
well within the domain where $J_0 > m_0$ already for a moderate
value of $J_0$, a large number of time steps is needed in order to
reach the asymptotic behavior and even longer times are necessary
for larger values of $J_0$ and/or smaller values of $T$. In
contrast, the EO procedure yields inconclusive results within
reasonable computing time, that could lead to wrong conclusions.
With the exception of the dip around $t \simeq 1575$, which is an
indication of crossover behavior in the overlap, the
consecutive-time correlation function $C_{t,t-1}$ is very close to
but not exactly equal to one through all the dynamics, which
indicates that there is almost always a very small fraction of
flipping spins, except at the crossover where that fraction is
about five percent over a few number of steps.

\begin{figure}[ht]
\center
\includegraphics[scale=0.70]{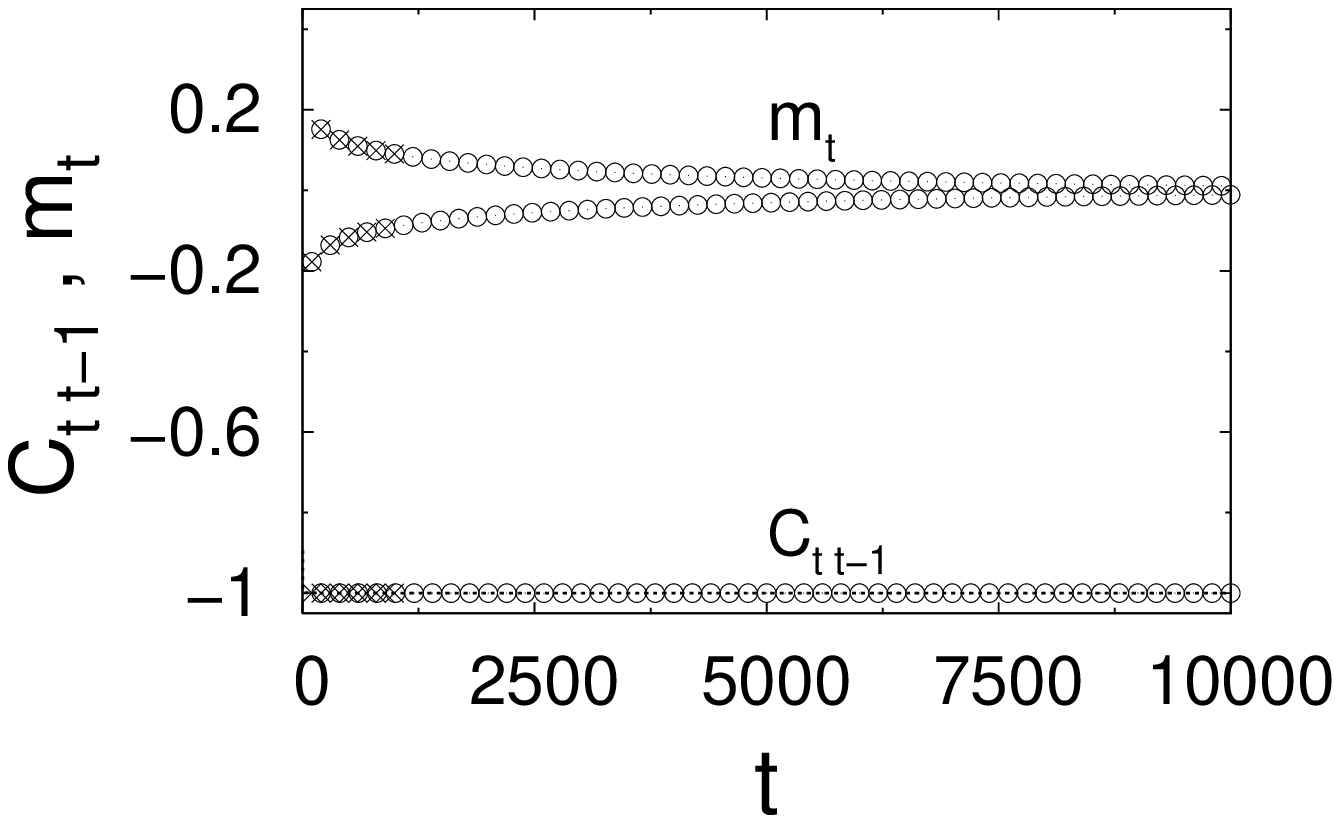}
\caption{Overlap and two-time correlation function for $T=0.08$,
$J_{0}=-0.5$, $m_{0}=0.4$ and $\alpha=0$. Superimposed are the
results with the EO procedure (dark circles).}
\label{dinamalpha0ciclo}
\end{figure}

We consider now the instability to synaptic noise of the frozen-in
cycles of period two in phase P, when $\alpha=0$. The results
obtained with the alternative approach and, for comparison with
the EO procedure, are shown in figure \ref{dinamalpha0ciclo} for
$J_0=-0.5$, $T=0.08$ and initial overlap $m_0=0.4$. The
oscillating overlap decreases continuously to zero with increasing
$t$, now with the parameters in the domain where $|J_{0}|> m_{0}$
and $J_0<0$ and, again, a large number of time steps is needed in
order to reach a vanishing overlap characteristic of a
paramagnetic phase. The two-time correlation function $C_{t,t-1}$
is very slightly larger than $-1$ through all the dynamics, which
indicates the permanent presence of a small fraction of units that
remains frozen between two consecutive times, causing a continuous
decrease in the amplitude of the cycles as time evolves. It turns
out that the decrease is slower for larger values of $|J_0|$
and/or smaller values of $T$. We remark, again, that the
convergence to the asymptotic behavior is rather slow and that
results based on a much shorter time scale could lead to the wrong
conclusion that there are cycles for finite $T$.

Consider next the situation for $|J_0|>|m_0|$ in the absence of
synaptic noise ($T=0$) and $\alpha\neq 0$. In this case we run into
difficulties with the EO procedure due to the vanishing denominator
in (\ref{16}) after a relatively small number of time-steps, and
the alternative dynamics turned out to be not a good approximation
in this situation. Nevertheless, one can still draw conclusions from
the calculations for a small number of time steps using the EO
procedure, as we show next.
\begin{figure}[ht]
\center \subfigure[$\,\,\alpha=0.005$ and $J_0=0.6$.]{
\includegraphics[scale=0.62]{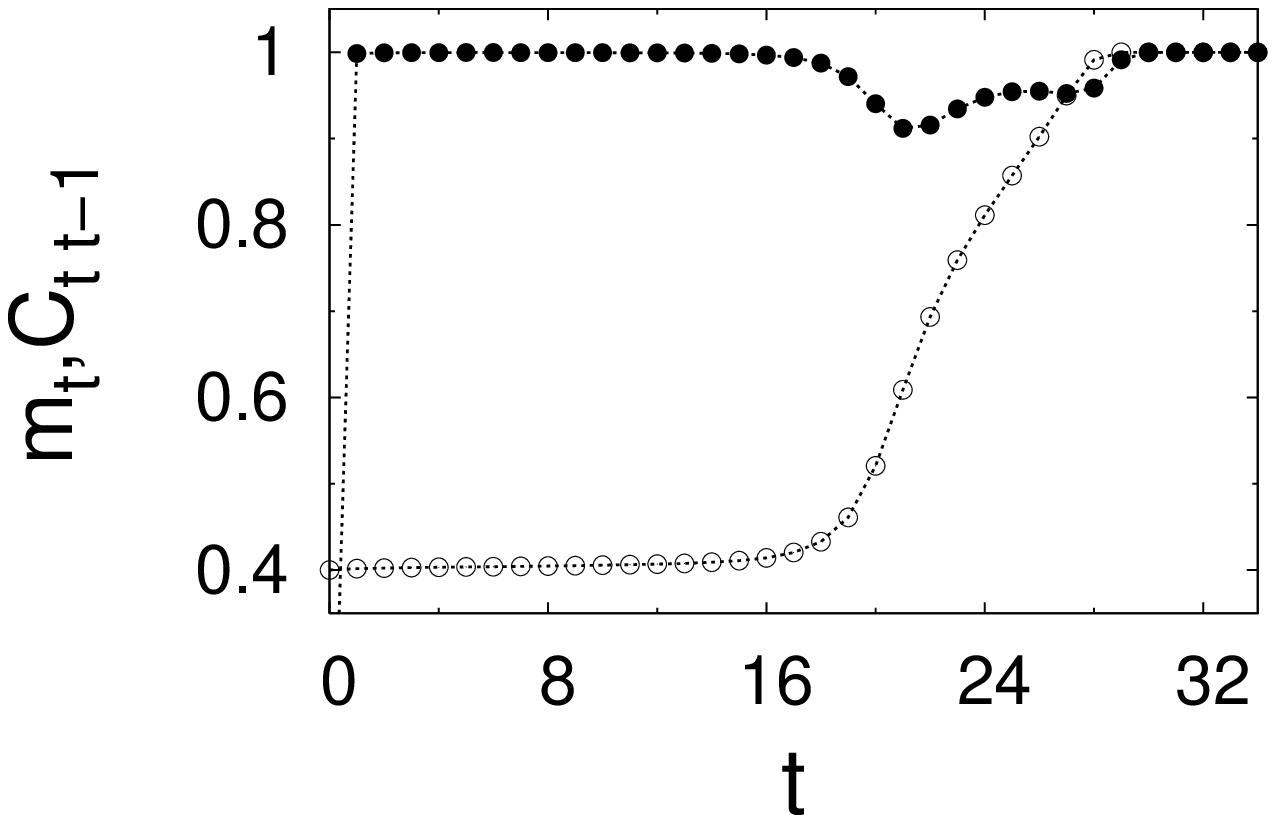}
\label{dinamalphafinretra} } \subfigure[$\,\,\alpha=0.04$ and $J_0
=-0.5$.]{
\includegraphics[scale=0.62]{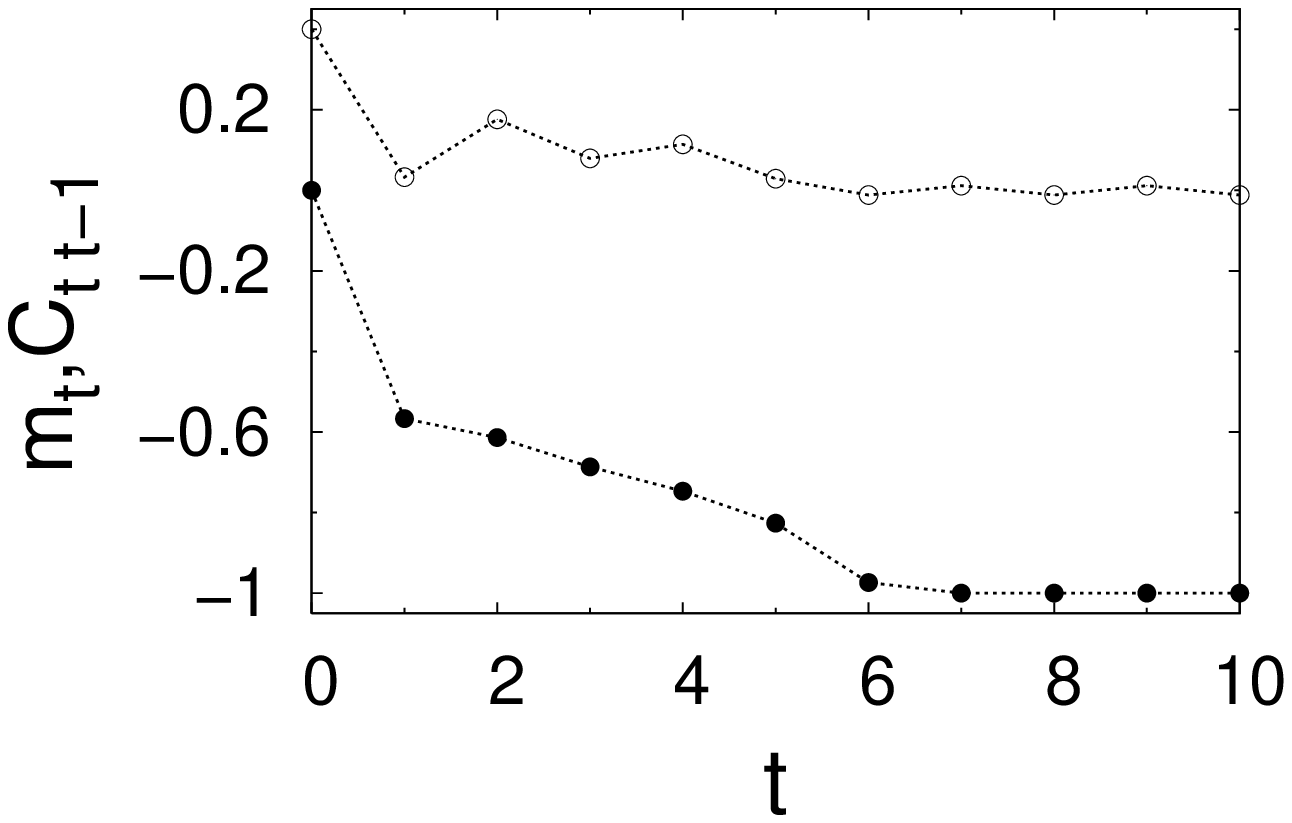}
\label{dinamalphafinretrb}
}
\caption{Overlap (open circles) and two-time correlation function
(full circles) at $T=0$, initial overlap $m_{0}=0.4$ and
$N_T=5 \times 10^{5}$ stochastic trajectories. The lines are a
guide to the eye.}
\end{figure}

In figure \ref{dinamalphafinretra} we illustrate the evolution of
the overlap and of the correlation function $C_{t,t-1}$ for
$J_0=0.6$, $\alpha=0.005$ and $m_{0}=0.4$, within the domain where
$J_0 > m_0$. There is a transient crossover from a frozen-in to a
retrieval state which takes place by means of an increase in the
fraction of flipping spins, shown by the dip in $C_{t,t-1}$,
similar to the behavior discussed above. The correlation function
between consecutive time steps is very close to but smaller than
one through all the dynamics, indicating a very small but finite
fraction of flipping spins outside the crossover region. It takes
a longer time interval to reach the crossover the bigger the value
of $J_0$ and/or the smaller the value of $\alpha$. For values of
$J_0$ closer to $m_0$ the transient crossover takes place for even
smaller values of $\alpha$, which suggests that this behavior
should occur for any $\alpha$. This, and the result obtained
above, illustrates the interesting feature of emergence of
retrieval behavior in the presence of a small amount of either
synaptic or stochastic noise, just enough to draw the network from
the frozen-in state.

For a larger $\alpha\simeq 0.3$ and the other parameters remaining
the same, we still find that $C_{t,t-1} < 1$ through all the
dynamics, but after an increase over the first few time steps the
overlap follows a very slow decrease as time evolves until a
remanent value is attained, which is characteristic of spin-glass
behavior \cite{SOK95}.

In figure \ref{dinamalphafinretrb} we exhibit the dynamical
behavior for $J_0=-0.5$, $\alpha=0.04$ and $m_{0}=0.4$, in which
the amplitude of the oscillating overlap decreases rapidly and
goes to zero. At the same time, the correlation function
$C_{t,t-1}$ decays to a value very close to $-1$ indicating an
almost full spin flip already after seven time steps. For smaller
values of $|J_0|$ and $\alpha$, within the domain where $|J_0| >
m_0$, a similar behavior is reached within a shorter number of
time steps. This behavior is reminiscent of the paramagnetic phase
in the presence of synaptic noise discussed above when $\alpha=0$.
A stationary state with vanishing overlap for $T=0$, $J_0 <0$ and
finite $\alpha$ is in qualitative agreement with a paramagnetic
phase found in the equilibrium replica symmetric approach
\cite{FK87}. It is worth noting that a paramagnetic state has also
been found in an asymmetric extremely diluted synchronous network
in the presence of stochastic noise at $T=0$ \cite{Fo88} and
$J_{0}<0$. We remind the reader, however, that we are dealing here
with a fully connected network with symmetric interactions.

In contrast, for a slightly larger value of $\alpha$, and the
other parameters being the same, $C_{t,t-1}$ no longer becomes
$-1$ and, instead, evolves towards a positive stationary value,
indicating a partial spin flip, together with an oscillating
overlap that decreases slowly until a finite remanent value is
reached, characteristic of spin-glass behavior.

Since our results suggest that either synaptic or stochastic noise,
with finite $T$ or $\alpha$ respectively, play a similar role in the
dynamics when $|J_0| > |m_0|$ in drawing the network from the
frozen-in states, one can expect that this will continue to be the
case in the presence of both kinds of noise. We checked this
explicitly for $m_0=0.4$, $T=0.08$, $\alpha=0.003$ and two values of
$J_0$. As expected, for $J_0=0.8$ we get the transient crossover
from a frozen-in state to retrieval behavior and, for $J_0=-0.5$, we
find a continuous decreasing amplitude of the oscillating overlap as
the time evolves, until a vanishingly overlap is reached. In both
cases we obtain $|C_{t,t-1}| < 1$ through the dynamics, indicating
either a small but finite fraction of flipping units or an almost
full spin flip, for $C_{t,t-1} > 0$ or $C_{t,t-1} < 0$,
respectively. Due to the small values of $T$ and $\alpha$, it is
necessary to go to larger times than those where the EO procedure is
practically applicable, and we resorted for this purpose to the
alternative approach which should be a good approximation for the
value of $\alpha$ used. In fact, for the smaller times, there is
very good agreement between the EO procedure with $N_T=5 \times
10^{5}$ and the alternative approach with $M=100$.

\section{Summary and conclusions}

We used the numerical simulation procedure of Eissfeller and Opper
and an alternative procedure developed in this work, both based on a
generating functional approach, in order to study the effects of
synaptic and stochastic noise on the dynamical evolution of Little's
model of a synchronous neural network with binary neurons and a
finite self-interaction $J_0$. We analyzed the stability to both
kinds of noise of frozen-in fixed points and frozen-in cycles that
occupy a large part of the space of parameters and that were found
in a previous work \cite{Fo88}. This is a crucial issue in the range
of parameters when there is a sizeable self-interaction such that
$|J_0|>|m_0|$, for which there is no guaranty that there will be a
flow to a retrieval fixed-point, even if $J_0 > 0$. That flow may be
expected only if $|J_0|<|m_0|$.

We found that the frozen-in states that appear at $T=0$ and
$\alpha=0$ are unstable to synaptic noise, with or without
stochastic noise, leading either to retrieval or to paramagnetic
states, for $J_0>0$ or $J_0<0$, respectively. Our work also
suggests the instability of the frozen-in states to stochastic
noise in the absence of synaptic noise, with the same kind of
final states. This implies the absence of period-two cycles in any
finite region of the phase diagram with $T$ and/or $\alpha$
different of zero.

One has to be certain that the true asymptotic state is reached in
the course of the dynamics. It is in order to deal with very long
transients that already appear for moderate values of $|J_0|$ and
low values of synaptic and/or stochastic noise, which are heavily
time consuming in the computations with the EO procedure, that we
developed our alternative dynamical procedure within the generating
functional approach. The procedure consists in neglecting the
retarded self-interaction term in the local field and performing
explicitly the summation over states that enters into the
calculation of the macroscopic quantities of interest, and it
amounts to the numerical solution of a set of equations for the
relevant macroscopic parameters that involves a Gaussian correlated
average, and it becomes exact in the absence of stochastic noise.
Thus, we can perform the computations over a much longer number of
time steps in a shorter computing time than the procedure of
Eissfeller and Opper. The procedure also yields results in very good
agreement with the EO procedure in the case of small stochastic
noise, finite $T$ and large values of $|J_{0}|$, within the time
scale where both procedures are applicable.

Finally, some concern about our alternative procedure. Although it
is exact only in the absence of stochastic noise, that is for
$\alpha=0$, our results for $T>0$ and a rather small $\alpha\neq 0$
should be reliable and it does not seem to be justified at this
stage, where we are not concerned with full phase diagrams, to
include memory effects in the retarded self-interaction due to the
states of the units at other than just the previous time step. On
the other hand, reasonable calculations can be done with the
alternative procedure over longer time scales which would be needed
to obtain results for larger ratios of $|J_0|/m_0$.

Despite the fact that the use of our alternative dynamics is
justified for sufficiently small stochastic noise (small values of
$\alpha$), it may be interesting to consider its extension for a
larger amount of noise. Even an approximate estimate of the
retarded self-interaction $R(t,t^{\prime})$, based on the
dominating features of the response function $G(t,t^{\prime})$,
would be useful for that purpose. A hint in that direction could
come from simulations of Eissfeller and Opper for
$G(t,t^{\prime})$. Preliminary calculations in that direction, for
small non-zero values of $T$ and $\alpha$, show that the response
function is close to zero for most times, except for the presence
of two peaks.

It is worth pointing out that the alternative procedure could be
useful in the dynamics of other synchronous neural network models
with a finite self-interaction $J_0$. The procedure should also be
useful for the study of the dynamics of other disordered systems.
Furthermore, the procedure can also be applied to neural network
models with $J_0=0$, including in the memory term of the effective
local field (\ref{11}) only the term which relates the state of
the system at time $t$ to that at time $t-2$. In this case, it
would be necessary to obtain an equation for the response
function. This, and related issues will be explored in a separate
work.

\ack

We thank J. F. Fontanari for an illuminating discussion. The work of
one of the authors (WKT) was financially supported, in part, by CNPq
(Conselho Nacional de Desenvolvimento Cient\'{\i}fico e
Tecnol\'ogico), Brazil. A grant from FAPERGS (Funda\c{c}\~ao de
Amparo \`a Pesquisa do Estado de Rio Grande do Sul), Brazil, to the
same author is gratefully acknowledged. F. L. Metz acknowledges a
fellowship from CNPq.

\end{document}